\documentclass[aps,prb,reprint,superscriptaddress]{revtex4-1}

\usepackage{CJK}
\usepackage{graphicx,amssymb,bm,amsfonts,amsmath,graphics,color,epstopdf}
\usepackage[bookmarks=false,pdfstartview=FitH,colorlinks=true,citecolor=blue,linkcolor=blue]{hyperref}
\newcommand {\bscco}{Bi$_2$Sr$_2$CaCu$_2$O$_{8+\delta}$}
\newcommand {\uJcm}{$\mu$J/cm$^2$}

\begin{document}
\begin{CJK*}{GBK}{}


\title{Time- and Momentum-resolved Gap Dynamics in \bscco}

\author{Christopher L.\ Smallwood}
\affiliation{Materials Sciences Division, Lawrence Berkeley National Laboratory, Berkeley, California 94720, USA}
\affiliation{Department of Physics, University of California, Berkeley, California 94720, USA}
\author{Wentao Zhang}
\affiliation{Materials Sciences Division, Lawrence Berkeley National Laboratory, Berkeley, California 94720, USA}
\author{Tristan L.\ Miller}
\affiliation{Materials Sciences Division, Lawrence Berkeley National Laboratory, Berkeley, California 94720, USA}
\affiliation{Department of Physics, University of California, Berkeley, California 94720, USA}
\author{Chris Jozwiak}
\affiliation{Advanced Light Source, Lawrence Berkeley National Laboratory, Berkeley, California 94720, USA}
\author{Hiroshi Eisaki}
\affiliation{Electronics and Photonics Research Institute, National Institute of Advanced Industrial Science and Technology, Ibaraki 305-8568, Japan}
\author{Dung-Hai Lee}
\affiliation{Materials Sciences Division, Lawrence Berkeley National Laboratory, Berkeley, California 94720, USA}
\affiliation{Department of Physics, University of California, Berkeley, California 94720, USA}
\author{Alessandra Lanzara}
\email[To whom correspondence should be addressed. Email: ]{alanzara@lbl.gov}
\affiliation{Materials Sciences Division, Lawrence Berkeley National Laboratory, Berkeley, California 94720, USA}
\affiliation{Department of Physics, University of California, Berkeley, California 94720, USA}
\date {\today}

\begin{abstract}
We use time- and angle-resolved photoemission spectroscopy to characterize the dynamics of the energy gap in superconducting \bscco\ (Bi2212). 
Photoexcitation drives the system into a nonequilibrium pseudogap state: Near the Brillouin zone diagonal (inside the normal-state Fermi arc), the gap completely closes for a pump fluence beyond $F\approx15$ \uJcm; toward the Brillouin zone face (outside the Fermi arc), it remains open to at least 24 \uJcm.
This strongly anisotropic gap response may indicate multiple competing ordering tendencies in Bi2212.
Despite these contrasts, the gap recovers with relatively momentum-independent dynamics at all probed momenta, which shows the persistent influence of superconductivity both inside and outside the Fermi arc.
\end{abstract}

\pacs{74.40.Gh, 74.25.Jb, 78.47.J-, 74.72.Kf}
\keywords{high-temperature superconductivity, cuprates, superconductivity, band gap, pseudogap, time-resolved ARPES, trARPES, pump-probe ARPES, angle-resolved photoemission spectroscopy, Bi2212, BSCCO}

\maketitle
\end{CJK*} 


%

\section{Introduction}

When a superconductor's electrons bind into Cooper pairs, they leave an energy gap in the electronic band structure that is strongly influenced by the strength, symmetry, and underlying character of the pairing mechanism within a given material.\cite{Tinkham} 
In high-temperature cuprate superconductors the pairing mechanism remains a matter of considerable debate, and there has been great interest in characterizing the details of both the superconducting gap and the possibly related pseudogap.
This latter gap exists in hole-doped cuprates at low carrier concentration near the Brillouin zone faces even above the superconducting critical temperature ($T_c$),\cite{Marshall96,Loeser96,Ding96} and leaves perplexing ungapped ``Fermi arcs'' near the Brillouin zone diagonals.\cite{Norman98} 

Several experimental studies have reported evidence that the superconducting gap and pseudogap are manifestations of intertwined yet separate charge ordering tendencies. 
For example, momentum-dependent gap measurements using angle-resolved photoemission spectroscopy (ARPES) indicate that the gap near the Brillouin zone face not only remains open above $T_c$, but exhibits peculiar structure at low and high temperature that is hard to explain in the context of superconductivity alone.\cite{Tanaka06,Hashimoto10,He11}
Specific evidence for a competing order, in the form of charge-density-wave stripes, has long been known to exist in La$_{2\!-\!x}$Sr$_x$CuO$_4$ and related lanthanum 214 compounds.\cite{Hayden92,Chen93,Tranquada94,Tranquada95,Fujita02,Kivelson03,Tranquada04} 
More recently, nuclear magnetic resonance\cite{Wu11} and x-ray scattering studies\cite{Ghiringhelli12,Chang12} have revealed that a charge density wave directly competes with superconductivity
in underdoped YBa$_2$Cu$_3$O$_{6+x}$.

In this letter, we use time-resolved ARPES to measure gap dynamics following the destruction of superconductivity in the cuprate \bscco\ (Bi2212) near optimal doping ($T_c=91$ K) by an ultrafast near-infrared laser pulse.
The study expands upon previous time-resolved ARPES works on cuprates\cite{Perfetti07,Graf11,Cortes11,Smallwood12,Zhang13,Rameau14} by providing gap measurements at a larger range of fluences than has previously been presented, by exploring a range of momentum space extending definitively beyond the normal-state Fermi arc, and by employing more advanced methods to characterize the nonequilibrium gap.

We report three primary findings. 
First, photoexcitation using a fluence (average optical energy deposited on a surface per unit area) greater than 15 \uJcm\ unambiguously drives the closure of the near-nodal gap, with a response time of 300--600 fs.
Because the gap is a direct manifestation of the superconducting order parameter, this result constitutes one of the most detailed characterizations to date of a nonequilibrium phase transition involving the destruction of superconductivity.
Second, we find significant momentum-dependent differences in gap sensitivity to photoexcitation: although the gap completely closes near the Brillouin zone diagonal, it remains open near the Brillouin zone face, establishing a transient pseudogap.
Such momentum-dependent differences support the existence of two (or more) competing orders in the cuprates.
Finally, we characterize gap recovery rates.
In spite of the nonequilibrium gap shift's amplitude variation, recovery rates throughout the probed crystal momentum range are nearly momentum-independent.
Thus, even in the presence of a competing-order scenario, the findings indicate that superconductivity continues to have a large influence on gap dynamics both inside and outside the Fermi arc region.

\section{Experimental details}

In a time-resolved ARPES experiment a crystalline sample is optically illuminated by a low-frequency pump pulse and an ultraviolet probe pulse in short succession.
The pump pulse drives the sample into a nonequilibrium electronic state, and the probe pulse initiates a photoemission event, ejecting electrons out of the sample where their momenta and energies can be measured. 
Nonequilibrium response dynamics are then characterized as a function of the time delay ($t$) between the pump and probe pulses.

Measurements in the present study were conducted using a hemispherical electron analyzer, and with pump and probe frequencies at 1.48 eV and 5.93 eV, respectively (see Ref.~\onlinecite{Smallwood12a} for complete details regarding the apparatus). 
The system's energy, momentum, and time resolutions are respectively 23 meV, 0.003 \AA$^{-1}$, and 300 fs. 
Data were acquired deep in the superconducting state at an equilibrium temperature of $T<20$ K, measured using a silicon diode placed in thermal contact with the sample.
The laser repetition rate was set to 543 kHz, ensuring that residual heating caused by the pump pulse was less than 20 K.
We corrected for detector nonlinearity following the prescription of Ref.~\onlinecite{Smallwood12a}.
Samples were grown using the traveling solvent floating zone method, and measured to be near optimal doping ($T_c=91$ K). 
In all cases, samples were cleaved in situ in a vacuum chamber maintained at pressures below $5\times10^{-11}$ Torr.

There is currently no established consensus on how best to characterize the nonequilibrium gap.
At equilibrium, the band gap appears as a feature in the single-particle spectral function, $A(\vec{k},\omega)$, which is given in general terms by the sum of two constituent parts, $A^-(\vec{k},\omega)$ and $A^+(\vec{k},\omega)$, corresponding to electron removal and electron addition. 
ARPES measures only $A^-(\vec{k},\omega)$ because electrons are extracted from the sample rather than added to it.\footnote{Inverse photoemission apparatuses exist, but at the time of this writing the energy resolutions of such systems exceed the magnitude of the superconducting gap in the cuprates, and no inverse photoemission experiment incorporates time-resolution.}
Progress in analyzing the gap can be achieved using the identity\cite{Randeria95}
\begin{equation}
A^-(\vec{k},\omega) = A(\vec{k},\omega)f(\omega), \label{product}
\end{equation}
which relates $A^-(\vec{k},\omega)$ to $A(\vec{k},\omega)$ through the Fermi-Dirac distribution function.

Out of equilibrium, $A^-(\vec{k},\omega)$ can be rigorously extended into the time-dependent form $A^-(\vec{k},\omega,t)$, provided that finite-duration pump and probe pulses are incorporated into the spectral function definition.\cite{Freericks09}
The relationship between $A^-(\vec{k},\omega,t)$ and the nonequilibrium gap is more complicated than a simple extension of Eq.~\ref{product} because the theoretical concept of temperature may no longer be well-defined.\cite{Moritz13,Sentef13}
Nevertheless, strategies borrowed from analyses commonly used at equilibrium can still provide insight.
In the following, we characterize the nonequilibrium gap in superconducting Bi2212 following two complementary techniques: (i) dividing the data by a Fermi-Dirac distribution function corresponding to an assumed electronic temperature $T_e(t)$, and (ii) symmetrizing time-dependent energy distribution curves (EDCs:\ ARPES intensity at fixed momentum) at the Fermi wave vector ($k_F$). 
Gap characterization using EDC symmetrization, which can be analyzed with smaller statistical uncertainty than the Fermi division analysis, is then carried out to extract detailed fluence and momentum-dependent gap recovery dynamics.

\section{Fermiology: Fermi-division analysis}

\begin{figure}[tbp]\centering\includegraphics[width=3.375in]{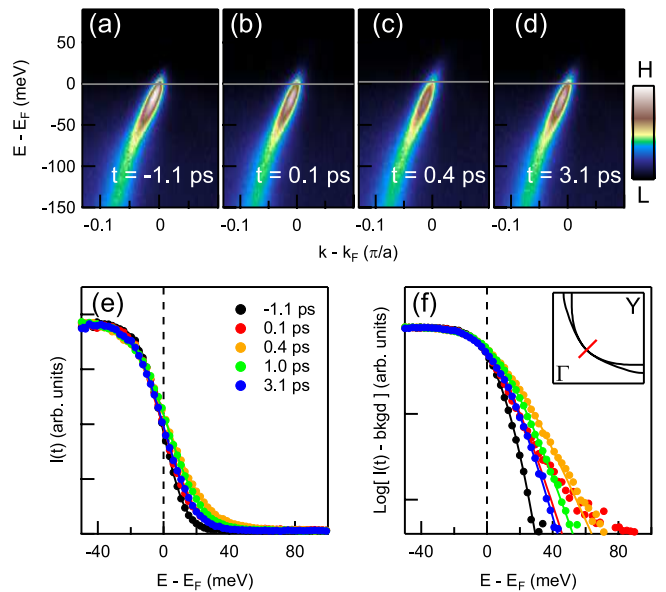}
\caption{\label{telec}Nodal quasiparticle relaxation dynamics, characterized using a $T_e$ and $\mu_e$ model.
{\bf(a)--(d)} Nodal time-resolved ARPES spectra at selected delay times.
{\bf(e)} Momentum-integrated EDCs (obtained by horizontally integrating the intensity for the spectra displayed in (a)--(d)) along with Fermi function fits at selected delay times. Solid curves are the result of fitting to Eq.~(\ref{fermifit}). 
{\bf(f)} Same data as in (e), but with the linear background above $E_F$ subtracted off, and displayed on a logarithmic scale to clarify the dynamics of quasiparticles far above the Fermi energy.
}
\end{figure}

To divide by an effective Fermi function, one must first extract a transient electronic temperature $T_e(t)$. 
This may be done by examining time-resolved spectra through a $k$-space cut intersecting one of the superconducting gap nodes, which occur along the Brillouin zone diagonals.
As shown in Fig.~\ref{telec}, we fit momentum-integrated EDCs for a cut along the $\Gamma$--$Y$ direction to the equation
\begin{equation}
I(E) = \left[ \frac{C_0 + C_1(E - \mu_e)}{\exp \left( \frac{E-\mu_e}{k_B T_e} \right) +1} +C_2 + C_3 E \right] * R(E), \label{fermifit}
\end{equation}
where $T_e$ and $\mu_e$ are fit parameters, $k_B$ is Boltzmann's constant, $R$ is a Gaussian resolution function of $\textrm{FWHM}=23$ meV, and the asterisk denotes convolution. 
The constants $C_0$, $C_1$, $C_2$, and $C_3$ allow the fit to account for an inelastic scattering background, for density-of-states variation, and for a linear background above $E_F$ caused by higher-order photoemission processes and spurious camera noise.

\begin{figure*}[tbp]\centering\includegraphics[width=7in]{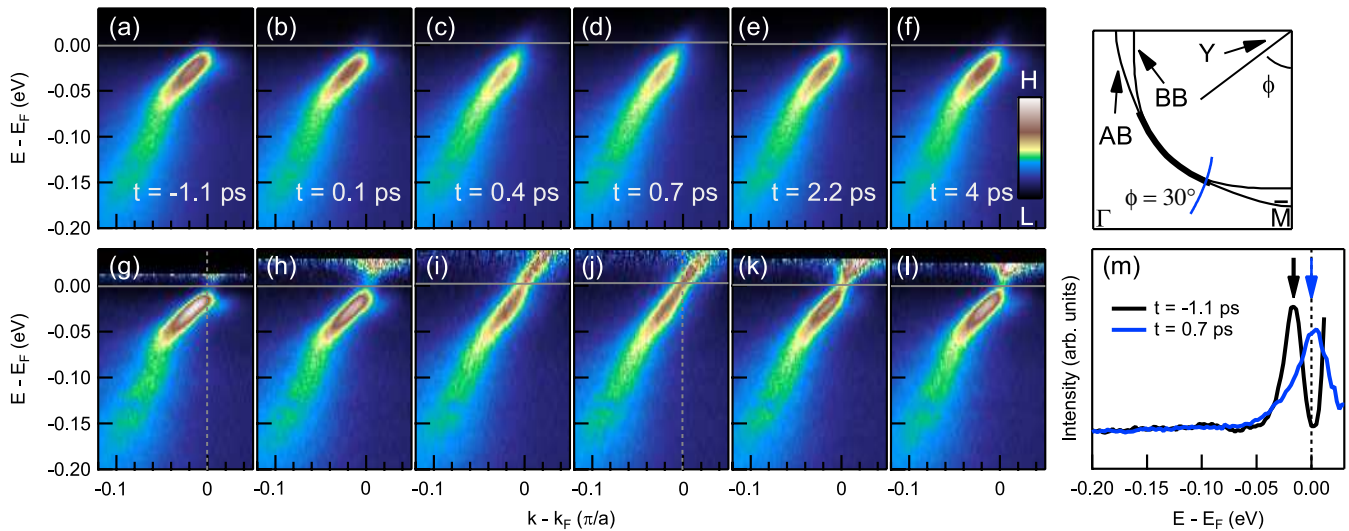}
\caption{\label{gapdiv}Near-nodal superconducting gap response to photoexcitation at a fluence of 23 \uJcm\ and an equilibrium temperature $T \ll T_c$.
{\bf(a)--(f)} Direct time-resolved ARPES intensity maps.
{\bf(g)--(l)} Intensity maps after applying a deconvolution procedure to remove the effect of the experimental resolution,\cite{Lucy74,Yang08} and dividing by an effective Fermi function.
{\bf(m)} EDCs at $k_F$ extracted from panels (g) and (j).
}
\end{figure*}

As already noted above, such an analysis requires an underlying assumption of a uniformly established electronic temperature, which must ultimately break down in Bi2212 for quasiparticles to be allowed to coherently oscillate with phonons,\cite{Kirchmann14} or to recombine with different rates at different points in $k$-space.\cite{Smallwood12}
However, there is precedent in the scientific literature for using a Fermi function to approximate the nonequilibrium distribution function in the cuprates,\cite{Perfetti07,Graf11} and a Fermi fit experimentally matches transient quasiparticles at the node with reasonable accuracy.
Fig.~\ref{telec} shows that although the fits and the data do not quite agree at the shortest times, they come into better agreement after 300 fs.

Along with the increase in $T_e$, there is also a slight pump-induced increase\cite{Rameau14} in the effective chemical potential $\mu_e$ (see the leading-edge shift between the data corresponding to $t=-1.1$ ps and $t=1$ ps in Fig.~\ref{telec}(e)). Further characterization of the band structure reveals that this is a rigid upward shift in the entire band, which may be caused by a transient change in the sample work function, by pump-induced space charge, or by the fact\cite{Ashcroft76} that an asymmetric density of states across $E_F$ can result in a mismatch between $\mu_e$ and $E_F = \lim_{T\to 0}\left[ \mu(T) \right]$.
The shift is small compared to the gap size and dynamics; it remains less than 4 meV for a fluence of 30 \uJcm, and vanishes to about 0.5 meV when the fluence is reduced to 4 \uJcm. None of the results reported here are affected by the shift.

\begin{figure}[tbp]\centering\includegraphics[width=3.375in]{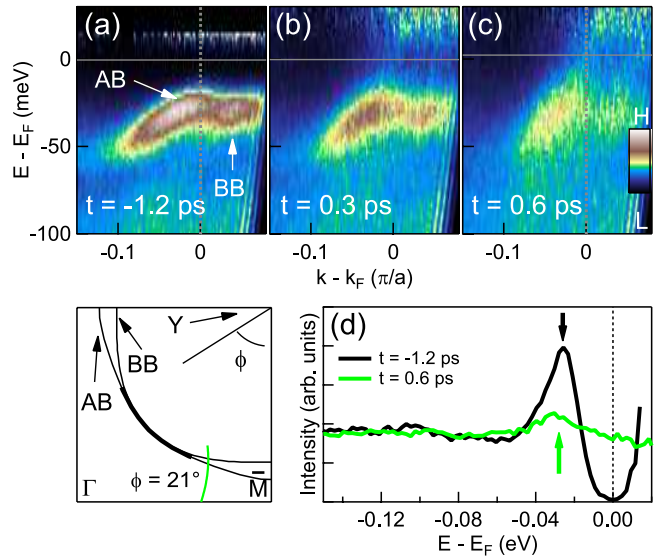}
\caption{\label{gapdiv2}Far-off-nodal superconducting gap response to photoexcitation at a fluence of 24 \uJcm\ and an equilibrium temperature $T \ll T_c$. As in Fig.~\ref{gapdiv}(g)--(l), the data have been subjected to a deconvolution procedure\cite{Lucy74,Yang08} and divided by an effective Fermi function.
{\bf(a)} Equilibrium spectrum. The magnitude of the energy gap is about 27 meV. Bilayer bonding bands (BB) and anti-bonding bands (AB) are visible.
{\bf(b)} and {\bf(c)} Transient spectra.
{\bf(d)} EDCs from panels (a) and (c) at the AB Fermi wave vector ($k_F$).
}
\end{figure}

\begin{figure*}[t]\centering\includegraphics[width=6.5in]{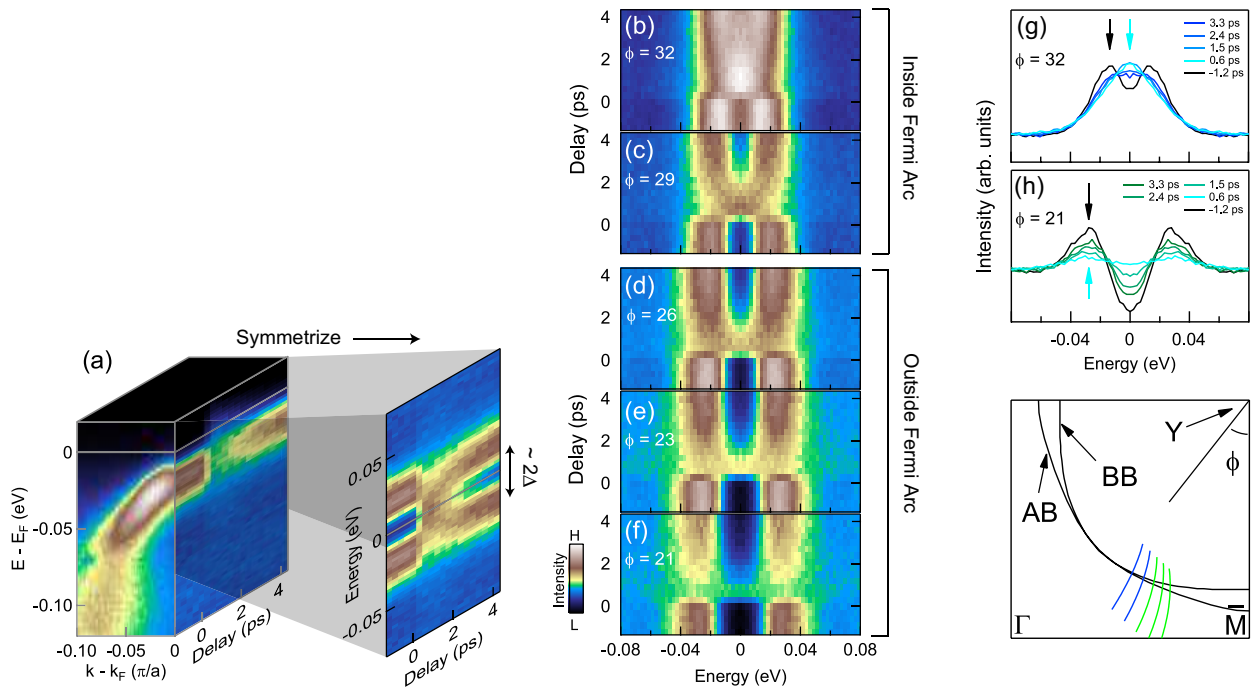}
\caption{\label{kdep}Momentum dependence of the transient superconducting gap for a pump fluence of 24 \uJcm\ and $T \ll T_c$, as analyzed using symmetrized energy distribution curves (EDCs).
{\bf(a)} Measurements are characterized based on EDCs at the AB Fermi wave vector ($k_F$), which are then symmetrized about the Fermi level to remove the effect of the electronic occupation function.\cite{Norman98a,Smallwood12}
{\bf(b)--(f)} False-color intensity plots of symmetrized EDC spectral weight versus energy and delay time. 
{\bf(g)--(h)}~EDCs at selected times for a representative near-nodal (g) and far-off-nodal (h) momentum cut. The black and cyan arrows highlight respective peak positions at $t=-1.2$ ps and $t=0.6$ ps.
}
\end{figure*}

Figures~\ref{gapdiv} and \ref{gapdiv2} show the results of the Fermi division analysis on two gapped cuts, corresponding respectively to $\phi=30^\circ$ and $\phi=21^\circ$ ($\phi$ is defined from the $Y$ point in $k$-space relative to $Y$--$\bar{M}$ as shown in the Fermi surface schematics).
Prior to dividing by the Fermi function, the data have also been numerically deconvolved along the energy dimension\footnote{Specifically, we applied 10 iterations of the Lucy-Richardson deconvolution algorithm.\cite{Lucy74,Yang08} Deconvolution was less effective when fewer than 10 iterations were used. Noise-induced artifacts became more prominent when more iterations were used.}
to mitigate the impact of finite energy resolution.
The equilibrium data, shown in Figs.~\ref{gapdiv}(a), \ref{gapdiv}(g), and \ref{gapdiv2}(a), exhibit several characteristic cuprate features, most prominently a gap magnitude that steadily increases between Brillouin zone diagonal and Brillouin zone face,\cite{Shen93} but also a well-defined dispersion kink at 70 meV that results from electron-boson coupling.\cite{Lanzara01}
Distinct bilayer bonding bands (BB) and anti-bonding bands (AB)\cite{Feng01,Chuang01a,Bogdanov01} are resolved in Fig.~\ref{gapdiv2}(a).

Nonequilibrium dynamics reveal that an infrared pump pulse of sufficiently high fluence dramatically affects the gap.
As shown in Fig.~\ref{gapdiv}, photoexcitation forces the near-nodal gap to completely close.
This can be seen both in the dispersion map shown in Fig.~\ref{gapdiv}(j), and in the comparison between equilibrium and transient EDCs shown in Fig.~\ref{gapdiv}(m).
The response of the gap accompanies a weakening of the 70 meV bosonic kink (compare panels (c) and (d) with panel (a)), which is discussed elsewhere.\cite{Zhang14}
Interestingly, the data also reveal that there is a slight delay between the application of the pump pulse and when the gap is maximally altered from its equilibrium state. 
After $t=2$ ps the gap reopens and the spectra begin to resemble those at equilibrium once again.

Figure~\ref{gapdiv2} shows a cut far from the node ($\phi=21^\circ$) and reveals that the gap response is highly anisotropic.
As with the near-nodal response, photoexcitation induces an increased ARPES intensity at the Fermi level, the response time is slightly delayed relative to the arrival of the pump pulse, and the dynamics are accompanied by a temporary weakening of the 70 meV bosonic kink.
However, the magnitude of the far-off-nodal gap, as reflected by the peak position of the lower Bogoliubov band, is only slightly shifted (see panels (c) and (d)).
As a result, the far-off-nodal gap response is more aptly characterized as filling in rather than closing.

Such dynamics could reflect a scenario where the far-off-nodal gap reflects a spatially integrated response of two coexisting forms of charge order, for example superconductivity and a competing pseudogap ordering tendency. 
Given that superconductivity is destroyed more easily by photoexcitation than the pseudogap, one would expect the superconducting component of the far-off-nodal gap to completely close in response to photoexcitation at 25 \uJcm. 
The pseudogap, meanwhile, remains open at the same fluence, and the resulting sum of the two signals would be a gap that appears to fill rather than to close.
Even if the far-off-nodal gap is exclusively governed by one ordering tendency, the fact that it fills rather than closes indicates that the origin of the far-off-nodal gap may be fundamentally distinct from the origin of the near-nodal gap.
A gap that fills without closing is characteristic of order being destroyed through phase fluctuations, for example, whereas a gap that closes is characteristic of the dynamics within a mean-field approximation like the Bardeen-Cooper-Schrieffer (BCS) model.\cite{Norman98a}

\section{Fermiology: Symmetrized EDC analysis}

The results are expanded with a more detailed momentum dependence of the gap in Fig.~\ref{kdep}, where EDCs at the AB Fermi wave vector ($k_F$) from several momentum cuts have been symmetrized.\cite{Norman98a,Smallwood12} 
The symmetrization procedure replaces the assumptions of a thermal analysis with an assumption of local particle-hole symmetry, and does not require line shape deconvolution,\cite{Norman98a} complementing the Fermi-division analysis.

Interestingly, the threshold between the dynamics of the near-nodal gap, which completely closes, and the far-off-nodal gap, which does not, occurs at $\phi=28^\circ$. 
This coincides with the momentum marking the end of the normal-state Fermi arc for optimally doped Bi2212 when $T$ is slightly greater than $T_c$ (based on synchrotron measurements\cite{Lee07} as well as equilibrium measurements taken using the present setup).
Inside the Fermi arc (at $\phi=32^\circ$ and $\phi=29^\circ$), the gap is fully closed by a fluence of 24 \uJcm, as can be seen in the false-color symmetrized EDC intensity plots in Fig.~\ref{kdep}(b)--\ref{kdep}(c).
The gap magnitude is less affected for cuts beyond the end of the Fermi arc ($\phi=26^\circ$, $\phi=23^\circ$, and $\phi=21^\circ$) at the same fluence. 
As shown in Fig.~\ref{kdep}(d)--\ref{kdep}(f), the gap remains open at all delay times although, as in Fig.~\ref{gapdiv2}(d) and \ref{gapdiv2}(e), there is an increased intensity at the Fermi level.
Such findings broadly support a coexisting-order scenario in the cuprates,\cite{Tanaka06,Lee07,Kondo09,He11} with the gap near the Brillouin zone diagonals predominantly reflecting superconductivity, and the gap near the Brillouin zone faces reflecting a distinct pseudogap order or combination of the pseudogap with superconductivity.
As in Figs.~\ref{gapdiv}(m) and \ref{gapdiv2}(d), the failure of the gap to completely close far away from the node is accompanied by an evolution from a gap that closes in response to photoexcitation to a gap that instead fills in. 
Figure~\ref{kdep}(g)--\ref{kdep}(h) shows selected symmetrized EDCs from cuts inside and outside the Fermi arc, which particularly highlight this difference.

We note that in the present study fluences beyond 25 \uJcm\ are not explored, and it is likely that the antinodal gap may be destroyed in addition to the near-nodal gap at even higher fluences.

\section{Quantitative dynamics}

\subsection{Near-nodal gap}

\begin{figure}\centering\includegraphics[width=3.375in]{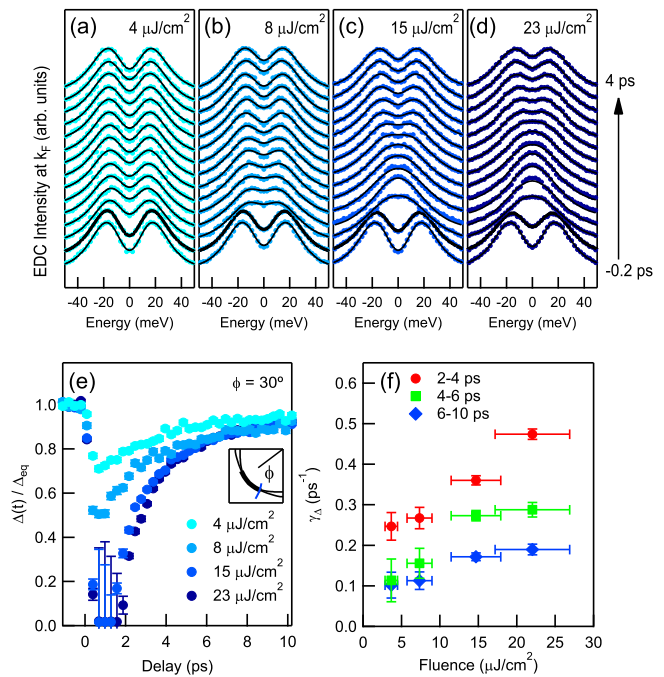}
\caption{\label{gapfluence}Fluence dependence of nonequilibrium gap dynamics inside the Fermi arc.
{\bf(a)--(d)} Symmetrized EDCs at $k_F$ and fit curves based on Eq.~(\ref{bcs}), for a gapped $k$-space cut at $\phi=30^\circ$ and $T=18$ K ($T\ll T_c$).
Bold curves correspond to $t=0$ ps.
{\bf(e)} Normalized gap magnitude versus pump-probe delay. 
{\bf(f)} Gap recovery rates $\gamma_\Delta$, extracted by fitting the data in (e) to Eq.~(\ref{gapfit}) between 2--4 ps, 4--6 ps, and 6--10 ps.
}
\end{figure}

In Fig.~\ref{gapfluence} we show the fluence dependence of the near-nodal nonequilibrium gap.
The delayed gap closure noted above is especially visible here and occurs at all fluences: the gap magnitude does not drop to its minimum until 300--600 fs after the application of the pump pulse.
A similar dynamic occurs in the nonequilibrium quasiparticle population,\cite{Smallwood12,Zhang13} and it is likely that the delays in the two phenomena are causally connected. 
Theoretical models of nonequilibrium superconductivity\cite{Owen72,Parker75} predict that an increased quasiparticle population should result in a decreased gap size if it helps the system's overall free energy achieve a minimum.
To characterize the gap quantitatively, we fit the data displayed in Fig.~\ref{gapfluence}(a)--\ref{gapfluence}(d) to the convolution of a Gaussian resolution function and the equation\cite{Norman98a}
\begin{equation}
I(\omega) = C_1 |\omega| + \frac{C_2 \Gamma}{(\omega - \Delta_k^2/\omega)^2+\Gamma^2}, \label{bcs}
\end{equation}
where $\Delta_k(t)$ corresponds to the energy of the gap, $\Gamma(t)$ corresponds to the peak width, and the leading term is added to account for the effects of an incoherent background.
Figure~\ref{gapfluence}(e) shows the trends in $\Delta_k(t)$, where it is clear that increasing the fluence beyond 15 \uJcm\ forces the near-nodal gap to completely close at 0.7 ps, in line with Figs.\ \ref{gapdiv}--\ref{kdep}.
The closure is closely affiliated with the destruction of superconductivity, and we note that the critical fluence here reported is in good agreement with an infrared pump and terahertz probe transmissivity study reporting that a fluence of 11 \uJcm\ results in a 90 percent loss of superfluid density.\cite{Carnahan04}
Features near 15 \uJcm\ have also been reported in the initial photoexcited quasiparticle population of optimally doped and underdoped samples of Bi2212.\cite{Zhang13}

In Fig.~\ref{gapfluence}(f) we characterize the fluence dependence of the near-nodal nonequilibrium gap recovery rate.
The gap recovery dynamics in Bi2212 are non-exponential.
However, meaningful trends in gap dynamics as a function of fluence and momentum can be extracted from the instantaneous gap recovery rate $\gamma_\Delta(t) \equiv \dot{\Delta}(t)/(\Delta(t) - \Delta_{eq})$, which can be obtained from exponential fits within short time intervals.
We extract $\gamma_\Delta$ by fitting $\Delta(t)$ to the function 
\begin{equation}
\frac{\Delta(t)}{\Delta_{eq}} = 1 - A_0 \, e^{-\gamma_{\Delta} (t-t_{\textrm{ref}})} \label{gapfit}
\end{equation}
between 2--4 ps, 4--6 ps, and 6--10 ps, as shown in Fig.~\ref{gapfluence}(f).
Such time intervals are chosen to be large enough to minimize statistical noise yet still small enough to return a reasonable goodness of fit.
In this equation, $\gamma_\Delta$ is the decay rate and $A_0$ is an amplitude defined at the freely selected time $t_{\textrm{ref}}$.
The recovery rate of the near-nodal gap is faster at higher fluences and shorter delay times.
These trends originate from a density-dependent response of the quasiparticle decay rate,\cite{Gedik04,Smallwood12} as well as from the aforementioned causal relationship between quasiparticle population and gap size, which is expected to be nonlinear. 
(At equilibrium, for example, the gap responds to the quasiparticle population according to the BCS gap equation.)

\subsection{Momentum dependent gap}

\begin{figure}\centering\includegraphics[width=3.375in]{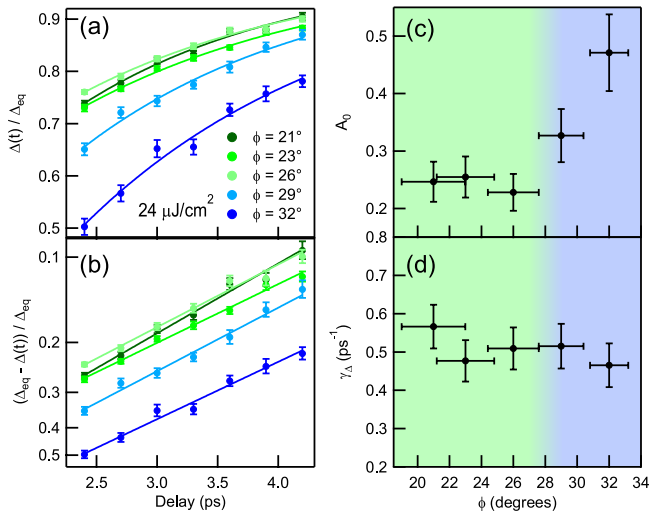}
\caption{\label{gaprates}Momentum dependence of the nonequilibrium gap with a pump fluence of 24 \uJcm.
{\bf(a)} Normalized gap magnitude vs.\ delay, and fits using Eq.~(\ref{gapfit}), where fits are extracted between 2.5 and 4 ps.
{\bf(b)} Data and fits from (a), shown on a logarithmic scale to highlight recovery rates.
{\bf(c)} Amplitudes ($A_0$) and {\bf(d)} recovery rates ($\gamma_\Delta$) corresponding to the gap magnitude shifts characterized in (a) and (b). $A_0$ is extracted at $t_{\textrm{ref}}\equiv 2.5$ ps.
}
\end{figure}

Figure~\ref{gaprates} shows an analysis of the momentum dependence of gap recovery rates between 2.4 and 4.2 ps for a fluence of 24 \uJcm, where $\gamma_\Delta$ is extracted from the data in Fig.~\ref{kdep} using Eqs.~(\ref{bcs}) and (\ref{gapfit}). 
The distinction between a gap that completely closes inside the Fermi arc and one that remains open outside the Fermi arc is clear in the amplitude dependence of the Eq.~\ref{gapfit} fit parameter $A_0$: as shown in Fig.~\ref{gaprates}(c), at 2.5 ps the gap measurements at $\phi=32^\circ$ and $\phi=29^\circ$ are suppressed by 47\% and 33\% of their equilibrium values, respectively, while the gap measurements at $\phi=26^\circ$, $\phi=23^\circ$ and $\phi=21^\circ$ are only suppressed by 20--25\% of their equilibrium values.
However, as shown in Figs.~\ref{gaprates}(b) and \ref{gaprates}(d), the nonequilibrium gap recovers with picosecond-scale dynamics at all probed momenta, and is independent of crystal momentum to within our uncertainty. 

\begin{figure}\centering\includegraphics[width=2.3in]{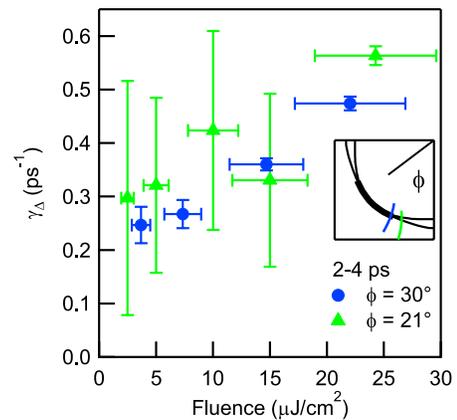}
\caption{\label{gaprates2}Fluence-dependent gap recovery rates for a representative cut inside ($\phi=30^\circ$) and outside ($
\phi=21^\circ$) the normal-state Fermi arc. Fits are extracted between 2 and 4 ps.
}
\end{figure}

Figure~\ref{gaprates2} shows a characterization of gap rates as a simultaneous function of momentum and fluence. 
Recovery rates far outside the Fermi arc are more difficult to characterize than those close to the node because of the very small gap amplitude shift. 
However, it is clear that the far-off-nodal gap recovery rate increases with increasing fluence just as it does close to the node, and near-nodal and far-off-nodal gap recovery rates are consistent with each other at all fluences. 
Previously, we reported a possible gap recovery rate dependence on momentum at low fluence, though the uncertainty in recovery rates was larger than the recovery rate difference.\cite{Smallwood12} 
The present study disconfirms a general trend of momentum-dependent gap dynamics as the pump fluence is increased to higher values.
These recovery rates should not be confused, however, with the dynamics of the quasiparticle population, where the recovery rates are clearly momentum-dependent.\cite{Smallwood12}

\begin{figure}\centering\includegraphics[width=2.5in]{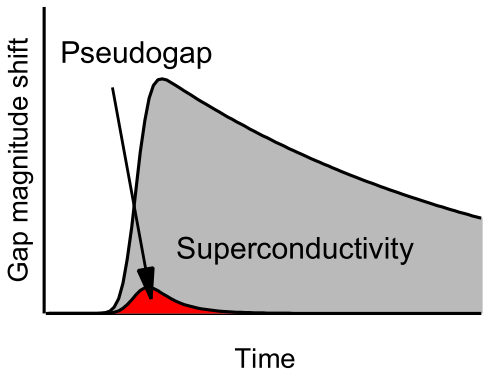}
\caption{\label{cartoon}Cartoon illustration of a possible mechanism for the relatively invariant gap recovery rate. If the component of the gap magnitude corresponding to superconductivity responds more strongly to photoexcitation than the component corresponding to the pseudogap, signatures of the latter signal may be washed out by the former signal.
}
\end{figure}

The uniformity in the gap recovery rates across the end of the normal-state Fermi arc is a puzzle, but could be explained by indications from studies using time-resolved reflectivity and transmissivity that the pseudogap state may recover as much as 10 times faster than superconductivity in Bi2212,\cite{Liu08a,toda11,Coslovich13} and by the fact that the magnitude of the pseudogap is largely unaffected by pumping in this fluence regime (i.e., the fractional gap magnitude shift outside the Fermi arc is small). 
Under these circumstances, distinct superconducting and pseudogap order parameters influence the equilibrium gap, but superconductivity dominates the gap dynamics. 
A cartoon depiction of the scenario is shown in Fig.~\ref{cartoon}, where the magnitude shift of the order parameters corresponding to the superconducting gap and a potentially competing pseudogap order (for example, a charge density wave) are displayed as a function of delay time. 
If superconductivity is more strongly affected by pumping than the competing order, then the signal due to superconductivity will dominate at even short times. 
At longer times the signal due to the competing order will be completely undetectable due to its faster recovery rate.
Such a two-order-parameter scenario is generally in agreement with equilibrium ARPES measurements reporting two distinct gaps in the cuprates if the crossover between a nodal superconducting gap and anti-nodal pseudogap occurs smoothly in $k$-space,\cite{Lee07} if both superconducting and pseudogap phenomena appear on equal footing at the antinode,\cite{He11} or if superconducting and pseudogap order parameters coexist at all momenta but are spatially separated in real space.

Momentum-independent $\gamma_\Delta$ values are also consistent with an alternate scenario suggested by gap studies using scanning tunneling spectroscopy, which postulates that the pseudogap just beyond the end of the Fermi arc is not the result of a competing order at all, but is rather a manifestation of phase-incoherent superconductivity.\cite{Lee09} 
Though phase competition is still predicted to exist, the onset of the competing order appears not at the end of the Fermi arc, but across the intersection of the normal-state Fermi surface with the antiferromagnetic zone boundary.\cite{Kohsaka08}
This intersection is beyond the present study's range of accessible momenta, but would be interesting to probe using time-resolved ARPES in future studies at higher probe photon energy. 
Regardless of the details, it is clear under both this and the previous scenario that superconductivity plays an important role in influencing quasiparticle dynamics both inside and outside the end of the Fermi arc.

Finally, the momentum-independent gap recovery rates could indicate that thermalization between quasiparticles at different momenta occurs rapidly in the high fluence regime. 
The gap recovery at all momenta would be governed by a common thermal order parameter under this scenario, which gains support from the fact that the distribution of nonequilibrium quasiparticles along the node resembles a thermal distribution to a significant extent for $t>300$ fs, and the success with which gap dynamics can be extracted given the quasi-thermal analysis employed in Figs.~\ref{telec}--\ref{gapdiv2}. 
However, we note that it is hard to reconcile thermal dynamics with the fluence and momentum dependencies previously reported in Bi2212 at lower fluence.\cite{Smallwood12} 

\section{Conclusions}

In conclusion, we have established that infrared photoexcitation using a pump fluence beyond 15 \uJcm\ definitively closes the superconducting gap near the Brillouin zone diagonals, that the gap remains open beyond the end of the normal-state Fermi arc up to at least 25 \uJcm, and that the gap recovers with nearly momentum-independent dynamics out to a Fermi surface angle of $\phi=21^\circ$ with recovery timescales on the order of picoseconds. 
We note that temporal onset dynamics associated with the complete quenching of the near-nodal gap provide an important benchmark for comparison with the optically induced destruction of other forms of order in strongly correlated materials, including magnetism\cite{Beaurepaire1996} and charge-density-wave order.\cite{Rohwer11} 
Beyond this, the results presented in this study have implications in the study of competing interactions in the cuprates more generally. 
For example, in demonstrating that photoexcitation induces a transient pseudogap, the results both add to a mounting set of experiments conducted at equilibrium supporting the existence of multiple competing phases in the cuprates, and they provide a complementary reference for ultrafast studies reporting evidence of the pseudogap in the nonequilibrium change in optical reflectivity. 
We hope that the dynamics and recovery trends here reported will stimulate many further discussions in the growing field of ultrafast phenomena in correlated systems.

\begin{acknowledgments}
We thank R.\ A.\ Kaindl, J.\ Orenstein, A.\ Vishwanath, G.\ Affeldt, and A.\ Fero for useful discussions.
This work was supported as part of the Quantum Materials Program at Lawrence Berkeley National Laboratory, funded by the U.S.\ Department of Energy, Office of Science, Office of Basic Energy Sciences, Materials Sciences and Engineering Division, under Contract No.\ DE-AC02-05CH11231.
\end{acknowledgments}

%

\end{document}